\documentstyle[sprocl,epsfig]{article}

\def\be{\begin{equation}}
\def\ee{\end{equation}}
\def\bea{\begin{eqnarray}}
\def\eea{\end{eqnarray}}
\def\nc{N_{\rm c}}
\def\nb{N_{\rm B}}
\def\E{{\bf E}}
\def\B{{\bf B}}
\def\j{{\bf j}}
\def\v{{\bf v}}
\def\x{{\bf x}}
\def\y{{\bf y}}
\def\l{l_{\rm mag}}
\def\md{m_{\rm D}}
\def\FFDUAL{F^{\mu \nu}\tilde{F}_{\mu \nu}}
\def\NCS{N_{\rm CS}}
\def\alphaw{\alpha_{\rm w}}
\def\gsim{\mbox{~{\raisebox{0.4ex}{$>$}}\hspace{-1.1em}
	{\raisebox{-0.6ex}{$\sim$}}~}}

\def\squeeze{\vspace{-0.03in}}

\begin{document}

\title{Do We Understand the Sphaleron Rate?\footnote
    {Talk presented at ``Strong and Electroweak Matter 2000'', 
     Marseille, France, June 14-17, 2000}
}

\author{Guy D. Moore}

\address
    {%
    Department of Physics,
    University of Washington,
    Seattle, Washington 98195
    }%


\maketitle\abstracts
{
I begin by answering a different question, ``Do we know the sphaleron
rate?'' and conclude that we do.  Then I discuss a crude but purely
analytic picture which provides an estimate of the sphaleron rate within 
the context of B\"{o}deker's effective theory.  The estimate, which
comes surprisingly close to the numerically determined sphaleron rate,
gives a physical picture of baryon number violation in the hot phase,
and provides a conjecture of the $\nc$ dependence of the sphaleron rate
in SU($\nc$) gauge theory.
}

\section{Introduction}

Originally I intended to entitle this talk, ``Do we Know the Sphaleron
Rate?''  Unfortunately the talk would then have been much too short:
$$
\mbox{\large Yes.}
$$

Speaking seriously, it is appropriate here to remind the audience of
what the 
sphaleron rate is, why we are interested in it, and how we go about
calculating it.  As I will discuss, we know the correct effective theory 
for determining the sphaleron rate; it is called B\"{o}deker's effective 
theory.  We also understand what the physics which goes into deriving
that effective theory, is.  Further, we can compute within the effective 
theory quite accurately.  But exactly what is going on, {\em within} the 
effective theory, is  more opaque.  In the second half of the talk I
will try to present my picture of what the physics is, which will lead
to a purely analytic estimate for the sphaleron rate 
which is closer than you might think to the right answer.

Baryon number (number of baryons minus anti-baryons, essentially the net 
amount of matter) is known to be very nearly conserved, but {\em not}
exactly conserved, within the standard model and under ordinary
conditions, as shown long ago by t'Hooft\cite{tHooft}.  The reason is
the anomaly, which shows that the baryon number current is coupled to
the SU(2) nonabelian field strength;
\be
\partial_\mu J_{\rm B}^\mu = N_{\rm F} \,
	\frac{g^2 \FFDUAL} {32 \pi^2} \, ,
\label{eq1}
\ee
where $F^{\mu \nu}$ is the SU(2) weak field strength and
$N_{\rm F}=3$ is the number of generations (and I have dropped an
irrelevant term involving hypercharge field strength).  Further, the total
derivative on the right hand side, integrated over spacetime in a vacuum 
to vacuum process, need not be zero; the topology of the group SU(2)
shows that it can be any integer, and equals the instanton number of the 
vacuum to vacuum process.  Hence baryon number is violated, but in
vacuum to vacuum processes its violation only occurs nonperturbatively.
Since the weak coupling is actually small, $\alphaw \simeq 1/30$, 
the instanton suppression factor $\exp(-4\pi/\alphaw)$ is
enormous, and baryon number violation is of no consequence under normal
conditions.  

However, the very hot conditions relevant in the very early universe did 
not constitute ``normal conditions.''  At
temperatures in excess of the electroweak phase transition (or
crossover) temperature $T \sim 80$GeV, the population functions for
infrared electroweak gauge bosons are large, and nonperturbative physics 
can be unsuppressed\cite{Linde,GrossPisarskiYaffe}.  This means that
baryon number violation becomes efficient; in fact it is only
polynomially, not exponentially, suppressed by the size of the weak
coupling\cite{ArnoldMcLerran}.

Since the universe has a small, nonzero, and very interesting (to us!)
abundance of baryons, understanding any cosmological process which
changes baryon number seems well motivated, especially when it is associated
with physics which we all believe must be there (the standard model
gauge group).  So there has been significant interest in understanding
the efficiency of baryon number violation in the early universe.

The sphaleron rate, the topic of this talk, is defined to be the
diffusion constant for the quantity on the right hand side in
Eq.~(\ref{eq1}), 
\be
\label{def_Gamma}
\Gamma \equiv \lim_{V\rightarrow \infty} \: \lim_{t \rightarrow \infty}
	\frac{\left\langle \left( \int \! dt \int \! d^3 \x \, 
	\frac{g^2 \FFDUAL}{32 \pi^2} 
	\right)^{\! 2}\, \right\rangle}{Vt} = \frac{1}{9} \, 
	\Gamma_{\! \nb} \, ,
\ee
with $\Gamma_{\nb}$ the diffusion constant for baryon number.
This is related to how fast a baryon number excess will decay, by a
fluctuation dissipation relation.  There is a free energy cost
associated with having a net baryon number;
\be
F = \frac{13}{12} \, \frac{\nb^2}{V T^2} \, .
\ee
Averaging over possible $\nb$ with $\exp(-F/T)$ weight gives
\be
\langle \nb^2 \rangle = \frac{6V T^3}{13} \, ,
\ee
which must be sustained by a balance between baryon number diffusion and 
baryon number decay, resulting in 
\be
\frac{1}{\nb} \, \frac{d\nb}{dt} = \frac{39}{4} \, 
	\frac{\Gamma}{T^3} \, .
\ee
Hence, the sphaleron rate tells us how fast a baryon number abundance
will decay.  Understanding its size is the object of the rest of the
talk.  

\section{B\"{o}deker's effective theory}

The context where we want to compute the sphaleron rate is the hot
standard model above the temperature of ``electroweak symmetry
restoration,'' which to good accuracy means\cite{Bodek_higgs} that we
may work in Yang-Mills theory.  It has been known for some time that the 
infrared behavior of Yang-Mills theory, even at weak coupling $\alphaw
\ll 1$, is non-perturbative\cite{GrossPisarskiYaffe}.  The easiest way
to understand this is to consider the thermodynamics.  It is well known
\cite{FKRS} that the full path integral for the thermodynamics of
the standard model is approximated up to corrections suppressed by
powers of $\alphaw$ by a 3-dimensional path integral,
\bea
Z & = & \int  {\cal D} A_i \, {\cal D} \Phi \, \exp(-H/T) \, , \nonumber \\
H & = & \int d^3 x \left[ \frac{1}{4} F_{ij}^a F_{ij}^a + 
	(D_i \Phi)^\dagger D_i \Phi + m^2 \Phi^\dagger \Phi
	+ \lambda \! \left( \Phi^\dagger \Phi \right)^{\! 2} \right] \, .
\label{part_func}
\eea
It is also well known that this 3-dimensional theory is
super-renormalizable.  That means that it has a characteristic momentum 
scale, in this case $g^2 T$.  On scales very large compared to this, it
is weakly coupled; in fact it grows more weakly coupled as a {\em power} 
of scale (until the effective theory breaks down at $p \sim \pi T$,
where it is as weakly coupled as the underlying theory); conversely, as
we look at more infrared scales it becomes more strongly coupled, until
at the scale $g^2T$ its interactions are nonperturbative.  Crudely
speaking we could say that this occurs because the occupation number of
a Bosonic state with momentum $p$ is $\sim T/p$; and that the loop
counting parameter is enhanced by one power of the occupation number.
This also suggests a nice way of understanding the physics which occurs
at strong coupling; large occupation number is the classical field
correspondence limit, and Eq.~(\ref{part_func}) is the partition
function of the classical field theory at finite temperature.

\begin{figure}[t]
\centerline{\epsfxsize=2.5in\epsfbox{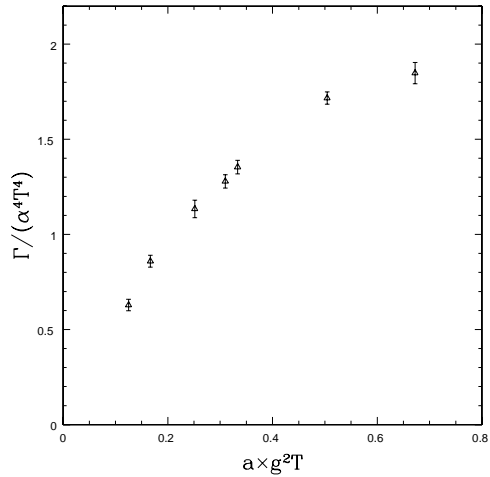}}

\caption{\label{pure_latt} $\Gamma$ for pure lattice classical 
Yang-Mills theory against lattice spacing.  It does not show rapid
convergence to a nonzero limit at zero spacing.  Data from 
\protect\cite{fine_latt}.}

\end{figure}

This correspondence led Grigoriev and Rubakov to conjecture
\cite{GrigRub} that the right effective theory to determine the
sphaleron rate is the classical field theory, regulated for instance on
a lattice.  We see that this cannot be true in Fig.~\ref{pure_latt}; the 
dependence on the lattice spacing is nontrivial and does not go away at
suitably small spacing.  This physics was predicted, before the data for 
the figure were taken, by Arnold, Son, and Yaffe\cite{ASY}.  
They argued correctly 
that ``hard'' (momentum scale $p \sim \pi T$) physics is important to
the dynamics of the infrared gauge fields, which are responsible for
Chern-Simons number diffusion.  Though not originally presented this
way, their argument boils down to fairly well known plasma physics.  A
plasma--abelian or nonabelian--is highly conducting.  In a conducting
medium, magnetic fields evolve slowly, because electric fields interact
strongly with the plasma.  In particular, the time evolution of a
magnetic field is related to the magnitude of an electric field by (the
nonabelian) Faraday's law.  But an electric field is quickly shorted
out, because it generates a current.  The nonabelian Ampere's law reads 
\be
- \dot{\E} + {\bf D \times} \B = \j \, .
\label{Ampere}
\ee

The greater the strength of screening in the medium, the greater the
importance of the current term, which because it is a conducting medium
is of form $\sim \int \E dt$, and causes the $\dot{\E}$ term to ``undo''
past electric fields.  More accurately, the current is
\be
\j^a(\x,t) = \md^2 \int \frac{d\Omega_v}{4\pi} \, \v \int \! \! dy \: 
	U^{ab}(\x,t;\x{-}\v y,t{-}y) \: \v{\cdot}\E^b(x{-}\v y,t{-}y) \, ,
\label{j_is}
\ee
with $\v$ a unit vector and its integral $d\Omega_v$, taken over the
unit sphere.  Here $U^{ab}$ is the straight line, adjoint parallel
transporter with initial and final points shown, which renders the
equation gauge covariant.  Together, Eq.~(\ref{Ampere}) and
Eq.~(\ref{j_is}) give classical Yang-Mills theory with added
hard thermal loop effects\cite{HTL,HTL2}.  (One must also add a noise
term, which has a form completely determined by equilibrium
thermodynamics; the noise correlator is nonlocal and we will not present 
it, see\cite{HuetSon,Son}.)  Arnold, Son, and Yaffe
advocated considering this effective theory, and argued that the
sphaleron rate should scale as $\Gamma \propto 1/\md^{2}$.

Two numerical implementations of this effective theory exist
\cite{particles,Wfields}.  Both are quite complicated and I will not
describe them here, though I will momentarily discuss their results for
$\Gamma$.  

B\"{o}deker has argued that it is possible to integrate out the scale
$gT$ from the effective theory just presented to arrive at a new, ``more 
effective'' theory\cite{Bodeker}.  The physical content of his
calculation (which is fairly complicated
\cite{Bodeker,Bodeker2,ASY2,LitimManuel}) is that scatterings between
the excitations carrying the current $j$, exchange the nonabelian
charges of the current carriers.  The mean rate for nonabelian charge
randomization turns out to be
\be
\gamma = \frac{\nc g^2 T}{4 \pi} \left[ \log \frac{\md}{g^2 T} + O(1)
	\right] \, .
\ee
This is parametrically larger than the scale where nonperturbative
physics happens, $\sim g^2T$, if we are willing to expand in
$\log(\md/g^2T)$.  Doing so, we conclude that a current carrier scatters 
frequently compared to the time it takes to traverse the length scale we 
are interested in, $\sim 1/g^2 T$.  Therefore, to leading order the
current can only depend on the local value of fields, and we may replace 
Eq.~(\ref{j_is}) with a local expression, which must be of form
\begin{equation}
\j = \sigma \E \, .
\end{equation}
For obvious reasons the constant $\sigma$ has been christened the
nonabelian conductivity [or ``color conductivity,'' if we think about
the SU($\nc$) theory rather than SU(2)].  It turns out one may determine 
$\sigma$ to next to leading log order; at this order it is
\be
\label{value_of_sigma}
\sigma^{-1} = \frac{3}{m_D^2} \, \gamma \, , \qquad
	\gamma = \frac{\nc g^2T}{4\pi} \left[ \ln \frac{m_D}{\gamma}
	+ 3.041 \right] \, .
\ee
The value for the constant $3.041$ is from \cite{AY2}.  Further, on the
scales of interest the term $\dot\E$ in Eq.~(\ref{Ampere}) can be
dropped relative to the current term, yielding B\"{o}deker's effective
theory, 
\be
\label{Bodek_theory}
\j = \sigma \E = {\bf D \times} \B + {\bf \xi} \, ,
\qquad
\left\langle \xi_i^a(\x,t) \xi_j^b(\y,t')\right\rangle = 2 \sigma T
	\delta^{ab} \delta_{ij} \delta(\x-\y) \delta(t-t') \, ,
\ee
with the noise Gaussian and white, with the autocorrelator shown.

\begin{figure}[t]
\centerline{\epsfxsize=3.0in\epsfbox{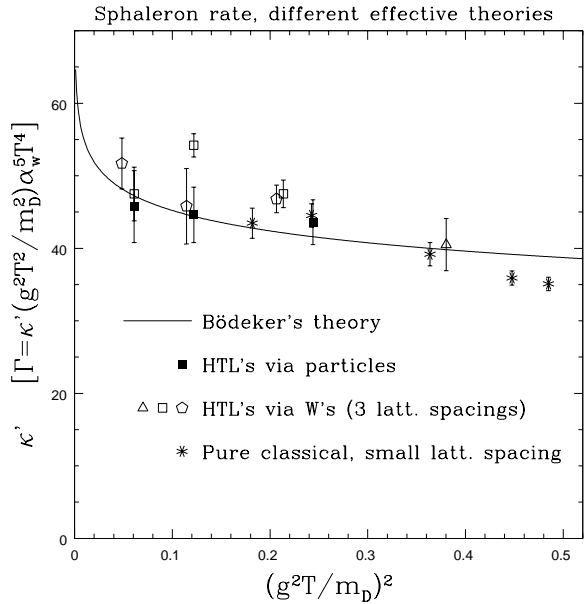}}
\squeeze
\caption{Sphaleron rate in B\"{o}deker's effective theory,
two lattice implementations of HTL effective theory
\protect\cite{particles,Wfields}, and pure lattice
theory interpreted as HTL effective theory (see
\protect\cite{Arnold_latt}). \label{compare}}
\squeeze
\end{figure}

To determine $\Gamma$ within this effective theory requires dealing with 
the nonlinear form of ${\bf D \times} \B$, necessarily including scales
where the appearance of the gauge field in ${\bf D}$ is as important as
the derivative term.  It is a hard nonlinear problem and the only
controllable techniques known are lattice techniques.  However as a
lattice problem it is particularly amenable to solution.  In particular, 
the effective theory has a good, well defined limit as the UV regulator
is taken to infinity\cite{ASY2}, and
it is possible to compute analytically a matching between a lattice
implementation and the continuum theory so the first errors are
quadratic in lattice spacing\cite{Bodek_paper}, and to define $\int
\FFDUAL$ topologically
\cite{slavepaper,broken_nonpert}.  One finds numerically
that\cite{Bodek_paper,Bodek_higgs}
\squeeze
\be
\Gamma = \kappa' \left( \frac{g^2 T^2}{\md^2} \right) \alphaw^5 T^4 \, , 
	\qquad
\kappa' = \Big( 10.0 \pm 0.2 \Big) \left[ \ln \frac{\md}{\gamma}
	+ 3.041 \right]\, .
\label{latt_result}
\ee
\squeeze
This is also in good agreement with the results of the HTL effective
theory, as summarized in Fig.~\ref{compare}.

\section{Model for the nonperturbative behavior}

I would argue that we understand pretty well {\em why} B\"{o}deker's
effective theory is the correct effective description of the the
infrared physics behind baryon number violation.  We understand why the
IR behavior is classical, and why the length scale $1/g^2T$ naturally
sets the scale for topology change.  We understand why IR gauge field
evolution is over-damped, in terms of plasma physics.  And we understand 
why, in a nonabelian theory, a nonabelian Ohm's law becomes valid on the
scale $1/g^2 T \log(1/g)$.  But the number in front of
Eq.~(\ref{latt_result}) was computed via lattice Monte-Carlo
means, which is sort of a ``black box.''  It would be nice to have some
physical picture of what is going on {\em within} the effective theory.
Here I provide a crude picture, which leads to a crude analytic estimate 
of the sphaleron rate in B\"{o}deker's effective theory.  

The point by point distribution of $\FFDUAL$ is as ultraviolet sick as
that of $F^2$.  So why is its diffusion constant finite and infrared
dominated?  The answer is that $\FFDUAL$ has topological meaning.  It is
a total derivative and can be written as the spacetime gradient of a
current, $(g^2/32\pi^2)\FFDUAL = \partial^\mu K_\mu$.  The associated
charge is the Chern-Simons number,
\be
\int d^3x K_0 \equiv \NCS = \frac{g^2}{32\pi^2} \! \int \! \! d^3x \;
	\epsilon_{ijk} \left( F^a_{ij} A_k^a - \frac{g}{3}
	f_{abc} A_i^a A_j^b A_k^c \right) \, .
\ee
In vacuum the integral of the $A^3$ term is always an integer; so there
must be nonvanishing fields for $\NCS$ to take on a non-integral value.
And in particular, since $F_{ij} A_k$ has one fewer derivative than
$F^2$, the more concentrated the region where $\NCS$ is stored, the
greater the energetic cost.  Physically, this means that a
fluctuation which creates a small amount of $\NCS$ leaves an imprint or
``strain'' 
behind in the fields, so there is gradient energy present until either
the fields relax in a way which undoes the $\NCS$ generated, or enough
more is added to get up to the next integer.  This is what prevents
there from being any UV divergences--in fact, any unsuppressed UV
contribution at all--in $\Gamma$.

The other piece of physics we know about the symmetric electroweak
phase, or pure 3-D Yang-Mills theory, is that there is magnetic
screening.  Colored degrees of freedom interact with each other
exclusively through gradient interactions, so a colored field at $\x$
feels the fields at $\y$ only through a parallel transporter from $\y$
to $\x$.  In fact, the field equations mean that the fields at $\x$ will 
generally evolve under the influence of $\y$ through some average over
paths for parallel transportation.  

For $\x$ and $\y$ of small separation $\ll 1/g^2 T$, two typical paths
between them will have approximately the same parallel transporter.
Therefore, the interactions via the different paths will add
coherently.  But for $\x$ and $\y$ separated by $\gsim 1/g^2 T$, a
generic pair of paths give completely different parallel
transportations.  (Equivalently, the mean trace of large
Wilson loops goes to zero.)  Roughly, there is a length scale $\l$,
called the magnetic screening length, beyond which parallel
transportation differs by $O(1)$ on change of path.  This means that the 
influence of fields at $\y$ on fields at $\x$ add incoherently over
paths and will approximately cancel.  A
field cannot ``see'' colored objects more than $\l$ away.  This arises
because of large nonabelian fields in between the two points, which are
responsible for making the parallel transportations differ.

The physical picture I argue for is that magnetic screening renders the
``gradient energy'' argument for keeping $\NCS$ from changing, ineffective for
field fluctuations on scales $\gsim \l$.  Hence, $\NCS$ {\em will}
diffuse due to $\FFDUAL$ from fields at the scale $l$.  These fields
will evolve diffusively; but fields on less IR scales will not
contribute to the diffusion of $\NCS$.

Within this picture, we can make a crude calculation of $\Gamma$, as
follows.  Consider some large volume $V$ as a collection of boxes $\l$
on a side.  Within each box, the most IR gauge field degrees of freedom
diffuse, and all higher wave number modes do not contribute; and the
contribution of different boxes are independent.  Writing $\FFDUAL = 4
E_i^a B_i^a$, Eq.~(\ref{def_Gamma}) becomes 
\be
\Gamma = \frac{1}{Vt} \sum_{\rm boxes} 
	\left\langle \left( \int dt' d^3x \frac{g^2
	E_i^a(x,t')B_{i,\rm IR}^a(x,t')}
	{8 \pi^2} \right)^{\! \! 2} \, \right\rangle\, ,
\ee
where the subscript on the magnetic field means that only the most
infrared component is kept.
The sum over directions gives the two
independent transverse polarizations, giving
\be
\frac{2 g^4}{64 \pi^4Vt}  \sum_{{\rm boxes}}
	\int \! dt_1 \, dt_2 \, d\x_1 \, d\x_2 \;
	\big\langle E^a(\x_1,t_1) \, E^b(\x_2,t_2) \, 
	B_{\rm IR}^a(\x_1,t_1) \, B_{\rm IR}^b(\x_2,t_2) \big\rangle \, .
\ee
We determine the $E$ field correlator from Eq.~(\ref{Bodek_theory}).
Our argument about scales says that, if we consider only the most
infrared component of the magnetic field, then we can neglect 
the gradient term, ${\bf D \times B}$, in
evaluating the $E$ correlators.  In this case $E$ and $B$ are
uncorrelated, and the $E$ correlators come only from the noise,
\be
\left\langle E^a(x_1,t_1) \, E^b(x_2,t_2) \right\rangle 
	= \frac{2T}{\sigma} \delta^{ab} \delta(x_1-x_2)
	\delta(t_1-t_2) \, .
\label{E_correlator}
\ee
(For higher frequency excitations the ${\bf D \times B}$ term is what
will ensure no contribution.)  
This performs one space and one time integral:
\bea
& & \int \!  dx_1 \, dx_2 \, dt_1 \, dt_2 \, \left\langle \, 
	E^a(x_1,t_1) \, E^b(x_2,t_2) \, 
	B_{\rm IR}^a(x_1,t_1) \, B_{\rm IR}^b(x_2,t_2) \, \right\rangle 
	\nonumber \\
& = & \frac{2T}{\sigma} \, \delta^{ab} \! 
	\int \! dx_1 \, dx_2 \, dt_1 \, dt_2 \, \delta(x_1{-}x_2)
	\, \delta(t_1{-}t_2) \,
	\left\langle B_{\rm IR}^a(x_1,t_1) \, B_{\rm IR}^b(x_2,t_2)  
	\right\rangle \nonumber \\
& = & \frac{2T}{\sigma} \, \delta^{ab} \! 
	\int_0^t \! dt_1 \int \! dx \, \left\langle B_{\rm IR}^a(x,t_1)
	\, B_{\rm IR}^b(x,t_1) \right\rangle \, .
\eea
This integral can be estimated by equipartition, 
\be
\int_0^t \! \!  dt_1 \! \int \! d^3x  \,
	\left\langle B_{\rm IR}^a(x,t_1) \, B_{\rm IR}^b(x,t_1) \right\rangle 
	= T \delta^{ab} \! \int_0^t \! dt_1 = Tt \, \delta^{ab} \, .
\ee
Probably there is more power in the infrared magnetic fields than
equipartition would estimate; but this is only one point at which our
analysis is crude.

The group indices have reduced to $\delta^{ab} \delta^{ab} = \nc^2{-}1$.
The volume of a box is $\l^3$, so in a total volume $V$ there are $V/\l^3$
boxes.  Putting everything together, we get
\be
\Gamma = \frac{4 (\nc^2{-}1) g^4 T^2}{64 \pi^4 \l^3 \sigma}
\, ,
\ee
which, on substituting Eq. (\ref{value_of_sigma}) and re-arranging, gives
\be
\Gamma = \frac{48 \nc^3(\nc^2{-}1)}{(\nc g^2T \l)^3} \left( 
	\frac{ \nc g^2T^2}{m_D^2}
	\right) \alphaw^5 T^4 \left[ \log
	\frac{\md}{\gamma} + 3.041 \right] \, .
\ee

It remains to estimate $\l$, which is a thermodynamic question and can
be answered within the 3-D Euclidean theory, Eq.~(\ref{part_func}).  
The only natural length
scale of hot Yang-Mills theory at weak coupling is $\l \sim 1/g^2T$, and
at large $\nc$ the $\nc$ dependence must be $\l \sim 1 / \nc g^2 T$.  The only
question is what the coefficient is.  I estimate it by trying to find on
what length scale the perturbative expansion for 3-D Yang-Mills theory
(describing the thermodynamics of the thermal theory) 
has completely broken down.  

One estimate is that
$\l \sim 1/p$ for the momentum $p$ with $|\Pi_{\rm T}(p)|=p^2$, 
that is, where the one loop transverse self-energy computed with tree
propagators is as large as the tree 
inverse propagator.  In Landau gauge  
(strict Coulomb gauge from the 3+1 dimensional perspective) this is
\be
\Pi(p,{\rm one \; loop, \: tree \; propagators}) 
	= \frac{11\nc g^2T}{64|p|} \left( \delta_{ij}p^2 - p_i p_j \right)
	\, ,
\ee
so $\l \sim 64/(11\nc g^2T)$.  This estimate is gauge dependent; in Feynman
gauge the 11 becomes 14.  However, since unequal separation correlators
are maximal in Landau gauge, it seems the appropriate gauge for
estimating where those correlators break down.

Alternately, we can take $\l$ as the
reciprocal of the ``magnetic mass'' found self-consistently by solving a
gap equation \cite{gap_guys}.  Without endorsing the efficacy of gap
resummed perturbation theory to make useful predictions, one can
nevertheless say that the mass found gives a characteristic
scale where higher order terms in the perturbative expansion are of
order the leading ones.  A two loop gap equation 
calculation \cite{new_gap_guys} gives $\l =
1 / m_{\rm mag} = 1 / 0.17 \nc g^2 T$ which agrees with the
self-energy estimate.\footnote{The 
computation \protect\cite{new_gap_guys} is in SU(2), 
but since all graphs at one and two loops
are planar, the $\nc$ dependence must be as shown.}

\begin{figure}[t]
\centerline{\epsfxsize=3.5in\epsfbox{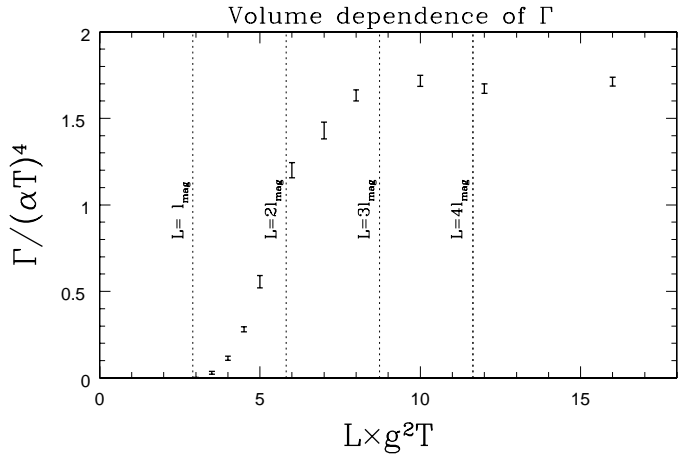}}
\caption{\label{vol_depend} Volume dependence of $\Gamma$ (pure lattice
theory, $g^2 aT=1/2$) illustrates that $\Gamma$ turns on for a box about 
$2\l$ across.}
\end{figure}

One check on this estimate of $\l$ is to look at lattice data in finite
volume, and see how large a lattice has to be before $\Gamma$ ``turns
on.''  We expect that it should have to be about $2\l$ across, because
we use periodic boundary conditions on the lattice.
Fig.~\ref{vol_depend} indicates that our estimate of $\l$ is
reasonable.  In particular, the inverse glueball mass $< 1/g^2T$ is {\em 
not} the appropriate scale.

Using the estimate for $\l$, we get
\be
\label{prediction}
\Gamma \sim 
	.24 \nc^3(\nc^2{-}1) \left( \frac{ \nc g^2T^2}{m_D^2} \right)
	\alphaw^5 T^4 \left[ \log \frac{\md}{\gamma} + 3.041 \right] \, .
\ee
Plugging in $\nc=2$
gives $\Gamma = 11.5 (g^2T^2/m_D^2) \alphaw^5T^4
\log(\ldots)$.  The 11.5 should be compared with
the actual number, $10.0$, presented earlier.  The quality of
the agreement is probably fortuitous.

\section{Conclusion}

After reviewing the current status of what the sphaleron rate {\em is},
and what the right effective theory for its determination is
(B\"{o}deker's effective theory), I have proposed a physical picture of
what is going on in the strongly coupled IR physics which determines the 
rate.  My picture is that, on scales longer than $\l$, magnetic
screening prevents ``any record of being kept'' of what past changes
have occurred to $\NCS$; therefore, very IR gauge fields evolve
diffusively.  I have presented a crude estimate of this diffusive
evolution and of $\l$, and find that the resulting sphaleron rate
$\Gamma$ is surprisingly close to the value determined in lattice
simulations.  

The model here should not be taken as an accurate quantitative predictor
of $\Gamma$; rather it should be viewed as a parametric
estimate where an effort has been made to include factors of 2 or $\pi$
where possible.  It also gives an interesting physical picture of $\NCS$
diffusion in Yang-Mills theory; rather than occasional, integer changes,
$\NCS$ diffuses from the incoherent accumulation of many contributions
individually much smaller than an integer.

A more robust prediction of this analysis is for the dependence of
$\Gamma$ on 
$\nc$.  Of course the scaling behavior $l \sim 1/\nc g^2T$ has $O(1/\nc^2)$
corrections, so the $\nc$ dependence in Eq. (\ref{prediction}) cannot
be relied on absolutely at small $\nc$. However, it would be interesting
to numerically investigate $\Gamma$ within B\"{o}deker's effective
theory for a few larger $\nc$ and see how
well the prediction holds.  In particular the case of SU(3) is
physically interesting\cite{strongsphal}.

In conclusion:  Do we understand the sphaleron rate?  I think, perhaps,
that we do.

\section*{Acknowledgments}
I thank Peter Arnold, Dietrich B\"{o}deker, Kari Rummukainen, 
Dam Son, and Larry Yaffe.
I particularly thank Larry Yaffe, who convinced me finally to present
these ideas.


\section*{References}
\begin{thebibliography}{99}
\bibitem{tHooft} G. t'Hooft, Phys. Rev. Lett. {\bf 37},8 (1976).




\bibitem{Linde}
A.~D.~Linde,
Phys.\ Lett.\  {\bf B96}, 289 (1980).

\bibitem{GrossPisarskiYaffe}
D.~J.~Gross, R.~D.~Pisarski and L.~G.~Yaffe,
Rev.\ Mod.\ Phys.\  {\bf 53}, 43 (1981).

\bibitem{ArnoldMcLerran} P. Arnold and L. McLerran, Phys. Rev. {\bf D 36},
        581 (1987).

\bibitem{Bodek_higgs}
G.~D.~Moore,
hep-ph/0001216.

\bibitem{FKRS}
K.~Farakos, K.~Kajantie, K.~Rummukainen and M.~Shaposhnikov,
Nucl.\ Phys.\  {\bf B425}, 67 (1994)
[hep-ph/9404201];
Nucl.\ Phys.\  {\bf B442}, 317 (1995)
[hep-lat/9412091].

\bibitem{GrigRub} D. Grigoriev and V. Rubakov, Nucl. Phys. {\bf B 299},
         248 (1988).

\bibitem{fine_latt}
G.~D.~Moore and K.~Rummukainen,
Phys.\ Rev.\  {\bf D61}, 105008 (2000)
[hep-ph/9906259].

\bibitem{ASY}
P.~Arnold, D.~Son and L.~G.~Yaffe,
Phys.\ Rev.\  {\bf D55}, 6264 (1997)
[hep-ph/9609481].

\bibitem{HTL} 
E. Braaten and R. Pisarski, Nucl. Phys. {\bf B337}, 569 (1990);
J. Frenkel and J. Taylor, Nucl. Phys. {\bf B334}, 199 (1990);
J. Taylor and S. Wong, Nucl. Phys. {\bf B346}, 115 (1990). 

\bibitem{HTL2}
J.P.~Blaizot and E.~Iancu,
Phys. Rev. Lett. {\bf 70}, 3376 (1993)
[hep-ph/9301236];
	V.P.~Nair,
	Phys. Rev. {\bf D48}, 3432 (1993)	
	[hep-ph/9307326];
	J.~Blaizot and E.~Iancu,
	Nucl. Phys. {\bf B421}, 565 (1994)
	[hep-ph/9401211];
	V.P.~Nair,
	Phys. Rev. {\bf D50}, 4201 (1994)
	[hep-th/9403146].

\bibitem{HuetSon} P.~Huet and D.T.~Son,
Phys. Lett. {\bf B393}, 94 (1997)
hep-ph/9610259].

\bibitem{Son} D.T.~Son,
UW/PT-97-19
[hep-ph/9707351].

\bibitem{particles}
G.~D.~Moore, C.~Hu and B.~Muller,
Phys.\ Rev.\  {\bf D58}, 045001 (1998)
[hep-ph/9710436].

\bibitem{Wfields}
D.~Bodeker, G.~D.~Moore and K.~Rummukainen,
Phys.\ Rev.\  {\bf D61}, 056003 (2000)
[hep-ph/9907545].

\bibitem{Bodeker}
D.~Bodeker,
Phys.\ Lett.\  {\bf B426}, 351 (1998)
[hep-ph/9801430].

\bibitem{Bodeker2}
.~Bodeker,
Nucl.\ Phys.\  {\bf B566} (2000) 402
[hep-ph/9903478];
Nucl.\ Phys.\  {\bf B559}, 502 (1999)
[hep-ph/9905239].

\bibitem{ASY2}
P.~Arnold, D.~T.~Son and L.~G.~Yaffe,
Phys.\ Rev.\  {\bf D59}, 105020 (1999)
[hep-ph/9810216];
Phys.\ Rev.\  {\bf D60}, 025007 (1999)
[hep-ph/9901304].

\bibitem{LitimManuel}
D.~F.~Litim and C.~Manuel,
Phys.\ Rev.\ Lett.\  {\bf 82}, 4981 (1999)
[hep-ph/9902430];
Nucl.\ Phys.\  {\bf B562}, 237 (1999)
[hep-ph/9906210];
Phys.\ Rev.\  {\bf D61} (2000) 125004
[hep-ph/9910348].

\bibitem{AY2}
P.~Arnold and L.~G.~Yaffe,
hep-ph/9912305;
hep-ph/9912306.

\bibitem{Bodek_paper}
G.~D.~Moore,
Nucl.\ Phys.\  {\bf B568}, 367 (2000)
[hep-ph/9810313].

\bibitem{slavepaper}
G.~D.~Moore and N.~Turok,
Phys.\ Rev.\  {\bf D56}, 6533 (1997)
[hep-ph/9703266].

\bibitem{broken_nonpert}
G.~D.~Moore,
Phys.\ Rev.\  {\bf D59}, 014503 (1999)
[hep-ph/9805264].

\bibitem{Arnold_latt}
P.~Arnold,
Phys.\ Rev.\  {\bf D55}, 7781 (1997)
[hep-ph/9701393].

\bibitem{gap_guys} W. Buchm\"{u}ller and O. Philipsen, Nuc. Phys. 
	{\bf B 443} (1995), 47;
	G. Alexanian and V. P. Nair, Phys. Lett. {\bf B 352} (1995),
	435; 
	R. Jackiw and S.Y. Pi, Phys. Lett. {\bf B 368} (1996), 131;
	J. M. Cornwall, Phys. Rev. {\bf D 57} (1998), 3694.
\bibitem{new_gap_guys} F. Eberlein, Phys. Lett. {\bf B 439} (1998), 130.
\bibitem{strongsphal} L. McLerran, E. Mottola, and M. Shaposhnikov,
        Phys. Rev. {\bf D 43} (1991) 2027;
	G. Giudice and M. Shaposhnikov, Phys. Lett.
        {\bf B 326} (1994) 118.

\end{thebibliography}
\end{document}